\documentclass[conference]{IEEEtran}
\pdfoutput=1 
\usepackage{hyperref}
\usepackage{color,overpic,cite}
\usepackage{xcolor}
\usepackage{pdfsync}
\usepackage{url}
\usepackage{amsmath,amsfonts,amssymb,amsthm,bbm,bm,comment,multirow,subfigure}
\usepackage{tcolorbox}
\usepackage[ruled,commentsnumbered]{algorithm2e} 

\newcommand{\eat}[1]{{}}

\DeclareMathOperator*{\argmax}{argmax}

\def\p{\widetilde}
\def\grad{\nabla}

\def\ll{\lambda}

\def\sumT{\sum_{t=1}^{T}}

\newtheorem{theorem}{Theorem}

\usepackage{enumitem}

\usepackage{mdframed}
\usepackage{bm}

\SetAlFnt{\small}
\IEEEoverridecommandlockouts
\title{Online Caching with Optimistic Learning} 

\begin{document}
\author{\IEEEauthorblockN{Naram Mhaisen\IEEEauthorrefmark{1}, George Iosifidis\IEEEauthorrefmark{1}, Douglas Leith\IEEEauthorrefmark{2}}%
	\vspace{2.5mm}
	\IEEEauthorblockA{
		\IEEEauthorrefmark{1}\textit{Software Technology Group, Delft University of Technology}, Netherlands.}
		\IEEEauthorrefmark{2}\textit{School of Computer Science and Statistics, Trinity College Dublin}, Ireland.}%
\maketitle

\begin{abstract}

The design of effective online caching policies is an increasingly important problem for content distribution networks, online social networks and edge computing services, among other areas. This paper proposes a new algorithmic toolbox for tackling this problem through the lens of \emph{optimistic} online learning. We build upon the Follow-the-Regularized-Leader (FTRL) framework which is developed further here to include predictions for the file requests, and we design online caching algorithms for bipartite networks with fixed-size caches or elastic leased caches subject to time-average budget constraints. The predictions are provided by a content recommendation system that influences the users viewing activity, and hence can naturally reduce the caching network's uncertainty about future requests. We prove that the proposed {optimistic} learning caching policies can achieve \emph{sub-zero} performance loss (regret) for perfect predictions, and maintain the best achievable regret bound $O(\sqrt T)$ even for arbitrary-bad predictions. The performance of the proposed algorithms is evaluated with detailed trace-driven numerical tests.

\end{abstract}

\section{Introduction}


\textbf{Motivation}. The quest for efficient data caching policies spans more than 50 years and remains today  one of the most important research areas for wireless and wired communication systems \cite{paschos-jsac}. Caching was first studied in computer systems where the aim was to decide which files to store in fast-accessible memory segments (\emph{paging}) \cite{Belady66}. Its scope was later expanded due to the explosion of Internet web traffic \cite{aggarwal-www} and the advent of content distribution networks (CDNs) \cite{RossComCom02}, and was recently revisited as a technique to improve the operation of wireless networks through edge caches \cite{femtocaching} and on-device caching \cite{femtocaching_d2d}. A common challenge in these systems is to design an online policy that decides which files to store at a cache, without knowing the future file requests, so as to maximize the cache \emph{hits} or some other more general cache-related performance metric.


There is a range of online caching policies that tackle this problem under different assumptions about the request arrivals. Policies such as the LFU and LRU are widely-deployed, yet their performance deteriorates when the file popularity is non-stationary, i.e., the requests are drawn from a time-varying probability distribution \cite{Sleator85, lru-sigmetrics08, lfu-sigmetrics99}. 
This motivated modeling non-stationary request patterns \cite{snm, kauffman} and optimizing accordingly the caching decisions \cite{mathieu, Elayoubi2015}. Another line of work relies on techniques such as reinforcement learning to estimate the request probabilities and make caching decisions  \cite{gunduz-reinforcement, giannakis-q-learning}; but typically these solutions do not scale nor offer optimality bounds. Caching was studied as an online learning problem in \cite{geulen2010regret, lykouris-ML} for a single-cache system; and in its more general form in \cite{paschos-infocom19} that proposed an online gradient descent (OGD) caching policy. Interesting follow-up works include sub-modular policies \cite{Li-online-2021}, online mirror-descent policies \cite{stratis-2020}, and the characterization of their performance limits \cite{abhishek-sigm20}. The advantage of these online learning-based caching policies is that they are scalable, do not require training data, and their performance bounds are \emph{robust} to any possible request pattern.

An aspect that has not been studied, however, is whether predictions about future requests can improve the performance of these learning-based caching policies. This is important in modern caching systems where most often the users receive content viewing recommendations from a recommendation system (\emph{rec-sys}). For instance, recommendations are a standard feature in streaming platforms such as YouTube and Netflix \cite{netflix}; but also in online social network platforms such as Facebook and Twitter, which moderate the users' viewing feeds \cite{recommend2}. Not surprisingly, the interplay between recommendations and caching has attracted substantial attention and recent works devised static joint  policies aiming, e.g., to increase the cache hit rate or reduce the routing costs by recommending to users already-cached files \cite{jordan-tmc, quek-twc21}. 

Changing vantage point, one can observe that since recommendations bias the users towards viewing certain content files, they can effectively serve as predictions of the forthcoming requests. This prediction information, if properly leveraged, can hugely improve the efficacy of caching policies, transforming their design from an online learning to an online optimization problem. Nevertheless, the caching policy needs to adapt to the accuracy of recommendations and the users propensity to follow them -- which is typically unknown and potentially time-varying. Otherwise, the caching performance might as well deteriorate by following these misleading request \emph{hints}. The goal of this work is to tackle exactly this challenging new problem and \emph{propose online learning-based caching policies which leverage predictions (of unknown quality) to achieve robust performance bounds}. 


\textbf{Contributions}. Our approach is based on the theory of Online Convex Optimization (OCO) that was introduced in \cite{zinkevich2003online} and has since been applied in different problems \cite{hazan-book}. The basic premise of OCO is that a learner (here the caching system)  selects in each slot $t$ a decision vector $x_t$ from a convex set $ \mathcal X$, without knowing the $t$-slot convex performance function $f_t(x)$, that change with time. The learner's goal is to minimize the growth rate of \emph{regret} $R_T\!=\!\sumT f_t(x^\star)\!-\!f_t(x_t)$, where $x^\star\!=\!\arg\max_{x\in \mathcal X} \sumT f_t(x)$ is the benchmark solution designed with hindsight. The online caching problem fits squarely in this setup, where $f_t(x)$ depends on the users requests and is unknown when the caching is decided. And previous works \cite{paschos-infocom19, abhishek-sigm20, stratis-2020} have proved that OCO-based policies achieve $R_T\!=\!O(\sqrt T)$, thus ensuring $\lim_{T\rightarrow \infty} R_T/T\!=\!0$. 

Different from these studies, we extend the learning model to include predictions that are available through the content recommendations. Improving the regret of learning policies via {predictions} is a relatively new area in machine learning research. For instance \cite{dekel-nips17} used predictions $\p c_t$ for the function gradient $c_t\!=\!\grad f_t(x_t)$ with guaranteed quality, i.e., $c_t^\top \p c_t \!\ge\! a \|c_t\|^2$, to reduce $R_T$ from $O(\sqrt T)$ to $O(\log T)$; and \cite{google-2020} enhanced this result by allowing some predictions to fail the quality condition. A different line of works uses regularizing functions which enable the learner to adapt to the predictions' quality \cite{sridharan-nips2013}, \cite{mohri-aistats2016}. This approach is more promising for the caching problem where the recommendations might be inaccurate, or followed by the users for only arbitrary time windows. 

Our approach relies on the Follow-The-Regularized-Leader (FTRL) algorithm \cite{shalev-ftrl} which we extend with predictions that offer \emph{optimism} by reducing the uncertainty about the next-slot functions. We first design a policy (OFTRL) for the bipartite caching model \cite{femtocaching}, which generalizes the standard single cache case. Theorem \ref{th:regret1} proves that $R_T$ is proportional to prediction errors ($\|c_t\!-\p c_t\|^2, \forall t$) diminishing to zero for perfect predictions; while still meeting the best achievable bound $O(\sqrt T)$ \cite{paschos-infocom19, abhishek-sigm20} even if all predictions fail. We continue with the \emph{elastic} caching problem \cite{akamai}, where the system resizes the caches at each slot based, e.g., on volatile storage leasing costs \cite{giannakis-elasticJSAC19, jungho-wiopt, akamai}. The aim is to maximize the performance subject to a long-term budget constraint. This places the problem in the realm of constrained-OCO \cite{giannakis-TSP17, paschos-icml, victor, johansson-TSP2020}. Using a new saddle point analysis with predictions, we devise Theorem \ref{th:regret-e} which reveals how $R_T^{(e)}$ and the budget violation $V_T^{(e)}$ depend on the caches and prediction errors, and how we can prioritize one metric over the other while achieving sublinear growth rates for both. 

The above algorithms make no assumption about the predictions accuracy, which might be high or low, or even alternate between these extremes (e.g., as user behavior changes) in any unpredictable and frequent fashion. However, in many cases, a rec-sys exhibits \emph{consistent} performance, namely its recommendations are of similar quality within a certain time window; either accurately due to recently trained model, or poorly due to e.g., distributional shift, see \cite{yang_recommender} and references therein. Our final contribution is a meta-learning caching framework that utilizes such consistent behavior in order to achieve \emph{negative} regret while maintaining sublinear regret when the consistency fails, see Theorem 3.


In summary, the contributions of this work are the following:

$\bullet$ Introduces an online learning framework for bipartite and elastic caching networks that leverages predictions to achieve a constant \emph{zero} regret for perfect recommendations and a sub-linear $O(\sqrt T)$ regret for arbitrary bad recommendations.

$\bullet$ Introduces a meta-learning framework that can achieve \emph{negative} regret by leveraging consistently-performing rec-sys.
	
$\bullet$ Evaluates the policies using various request models and real datasets \cite{zink2008watch} and compares them with key benchmarks.

The work presents conceptual innovations, i.e., using recommendations as predictions for caching, and using different online caching algorithms in a meta-learning algorithm; as well as technical contributions such as the new optimistic FTRL algorithm with budget constraints (Theorem  \ref{th:regret-e}). While we focus on data caching, the proposed algorithms can be directly applied to caching of services on edge systems.
\section{System model and Problem Statement}\label{sec:model}

\begin{figure}[!t]
	\centering
	\includegraphics[width=2.40in]{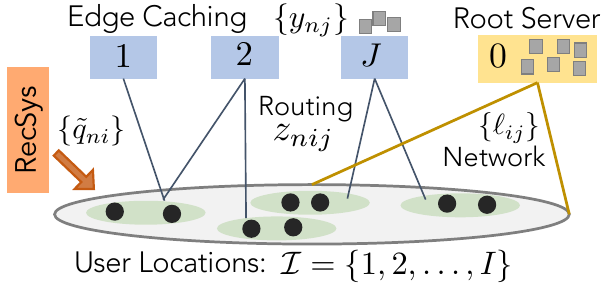} 
	\vspace{-1mm}
	\caption{\textbf{System Model}. A network of $\mathcal J$ caches serves file requests  from a set $\mathcal I$ of user locations. Unserved requests are routed to the Root Server. Caching decisions are aided via the recommendations provided by the  rec-sys.}
	\label{fig:system}
	\vspace{-10mm}
\end{figure}
\subsection{Model Preliminaries}

\textbf{Network}. The caching network includes a set of edge caches ${\cal J}\!=\!\{1,2,\dots, J\}$ and a root cache indexed with 0, Fig. \ref{fig:system}. The file requests emanate from a set of non-overlapping user locations ${\cal I}=\{1,2,\dots, I\}$. The connectivity between $\mathcal I$ and $\mathcal J$ is modeled with parameters $\ell\!=\!\big(\ell_{ij}\in \{0,1\}: i\!\in\!\mathcal{I}, j\!\in\!\mathcal{J} \big)$, where $\ell_{ij}\!=\!1$ if cache $j$ can be reached from location $i$. The root cache is within the range of all users in $\mathcal{I}$. This is a general non-capacitated bipartite model  \cite{paschos-book} that encompasses as a special case the celebrated femtocaching model \cite{femtocaching}, and can be used both for wired and wireless networks.

\textbf{Requests}. The system operation is time slotted, $t\!\!=\!\!1,2,\dots,\!T$. Users submit requests for obtaining files from a library $\mathcal{N}$ of $N$ files with unit size; we note that the analysis can be readily extended to files with different sizes. Parameter $q^{t}_{ni}\!\in\!\{0,1\}$ indicates the submission of a request for file $n \! \in \! \mathcal{N}$ by a user at location $i \! \in \!\mathcal{I}$ in the beginning of slot $t$. At each slot we assume there is one request; i.e., the caching decisions are updated after every request, as in LFU and LRU policies,  \cite{giovanidis-mLRU, leonardi-implicit}. Hence, the request process comprises successive vectors $q_t\!=\!(q_{ni}^t\!\in\!\{0,1\}: n\!\in\!\mathcal N, i\!\in\!\mathcal I)$ from the set:
\begin{equation}
	\mathcal{Q}=\bigg\{q\in \{0,1\}^{N\cdot I} ~\Big |~ \sum_{n\in\mathcal{N}}\sum_{i\in\mathcal{I}} q_{ni}=1\bigg\}. \notag
\end{equation}
We make no assumptions for the request pattern; it might follow a fixed or time-varying distribution that is unknown to the system; and can be even selected strategically  by an \emph{adversary} aiming to degrade the caching operation. If a policy performs satisfactory under this model, it is ensured to achieve (at least) the same performance  for other request models.

\textbf{Recommendations}. There is a recommender system (\emph{rec-sys}) that suggests up to $K_i$ files to each user $i\!\in\!\mathcal I$, see \cite{netflix} for the case of Netflix. User $i$ requests a recommended file with a certain probability that captures the user's propensity to follow one of the recommendations. Unlike prior works that consider these probabilities fixed \cite{akis-wowmom20, jordan-tmc}, we model them as unknown and possibly time-varying.
A key point in our approach is that the content recommendations, if properly leveraged, can serve as {predictions} for the next-slot user requests which are otherwise unknown. We denote with $\tilde q_t$ the prediction for the request $q_t$ that the system will receive at the beginning of slot $t$, and we assume that $\tilde q_t$ is available at the end of slot $t\!-\!1$, i.e., when the rec-sys provides its recommendations.




\textbf{Caching}. Each cache $j\!\in\!\mathcal J$ stores up to $C_j\!<\!N$ files, while the root cache stores the entire library, i.e., $C_0\!\geq \!N$. We also define $C\!=\! \max_{j\in\mathcal J} C_j$. Following the standard femtocaching model \cite{femtocaching}, we perform caching using the \emph{Maximum Distance Separable} (MDS) codes, where files are split into a fixed number of $F$ chunks, which include redundancy chunks. A user can decode the file if it receives any $F$-sized subset of its chunks. For large values of $F$, the MDS model allows us to use continuous caching variables.\footnote{Large files are composed of thousands of chunks and hence the rounding operation induces practically negligible errors \cite{paschos-book}} Hence, we define the  variable $y_{nj}^t\!\in\![0,1]$ which denotes the portion of $F$ chunks of file $n\!\in\!\cal N$ stored at cache $j\!\in\! \cal J$, and we introduce the $t$-slot caching vector $y_t\!=\!(y_{nj}^t: n\!\in\! \mathcal N, j\!\in\!\mathcal J)$ that belongs to set: 
\[
\mathcal{Y}=\bigg\{y\in [0,1]^{N\cdot J} ~\Big|~ \sum_{n\in\mathcal{N}}y_{nj}\leq C_j, ~j\in \mathcal J\bigg\}.
\]

\textbf{Routing}. Since each user location $i\in\mathcal{I}$ may be connected to multiple caches, we need to introduce  routing variables. Let $z_{nij}^t$ denote the portion of request $q_{ni}^{t}$ served by cache $j$. In the MDS caching model the requests can be simultaneously routed from multiple caches and, naturally, we restrict\footnote{This practical constraint is called the \emph{inelastic} model and compounds the problem, cf. \cite{abhishek-sigm20} for a comparison with the simpler elastic model.}  the amount of chunks not to exceed $F$. Hence, the $t$-slot routing vector $z_t=(z_{nij}^t\!\in\![0,1]: n\!\in\!\mathcal N, i\!\in\!\mathcal I, j\!\in\!\mathcal J)$ is drawn from:
\[
\mathcal{Z}=\bigg\{z\in [0,1]^{N\cdot J\cdot I} ~\Big|~ \sum_{j\in\mathcal{J}}z_{nij}\leq 1, ~n\in \mathcal N, i\in\mathcal I\bigg\}.
\]
Requests that are not (fully) served by the edge caches $\mathcal J$ are served by the root server that provides the missing chunks. This decision needs not to be explicitly modeled as it is directly determined by the routing vector $z_t$.

\subsection{Problem Statement} 

\textbf{Cache Utility \& Predictions}. We use parameters $w_{nij}\in[0, w]$ to model the system utility when delivering a chunk of file $n\!\in\! \cal N$ to location $i\!\in\! \cal I$ from cache $j\!\in\!\cal J$, instead of using the root server. This utility model can be used to capture bandwidth or delay savings, and other edge-caching gains in wired or wireless networks. The caching benefits can in general differ for each cache and user location, and may vary with time as it is explained in the sequel. Note that the cache-hit maximization problem is a special case of this more general setting \cite{paschos-jsac}. To streamline presentation we introduce vector $x_t\!= \!(y_t, z_t)\in \mathbf R^m$, with $m\!=\!NIJ\!+\!NJ$, and define the system utility in slot $t$ as:
\begin{equation}\label{eq:biput}
	f_t(x_t)= \sum_{n\in\mathcal{N}}\sum_{i\in\mathcal{I}}\sum_{j\in\mathcal{J}} w_{nij} q^t_{ni}z_{nij}^t\ ,
\end{equation}
and we denote its gradient $c_{t+1}\!=\!\nabla f_{t+1}(x_{t+1})$. As it will become clear, our analysis holds also for non-linear concave functions $f_t(x)$; this generalization is useful in case, e.g., we wish to enforce fairness in the dispersion of caching gains across the user locations \cite{jungho-wiopt}.

The main challenge in online caching is the following: at the end of each slot $t$ where we need to decide the cache configuration, the utility function $f_{t+1}$ is not available. Indeed, this function depends on the next-slot request $q_{t+1}$ that is revealed only after $y_{t+1}$ is fixed\footnote{In our case, since the routing is directly shaped by the caching, this restriction affects also $z_{t+1}$.}, see \cite{paschos-infocom19, abhishek-sigm20, lykouris-ML}. Besides, this is also the timing of the LRU/LFU policies \cite{giovanidis-mLRU, leonardi-implicit}. However, the recommendations provided to users can be used to form a predicted request $\p q_{t+1}$. For example, the caching system can set $\p q_{\hat n\hat i}^{t+1}\!=\!1$ and $\p q_{ni}^{t+1}\!=\!0, \forall (n,i)\neq (\hat n, \hat i)$, where $(\hat n, \hat i)$ is the request with the highest predicted probability\footnote{Note that our caching policy is orthogonal to the mechanism that maps the recommendations to predictions.}. Then, we can use $\p q_{t+1}$ to create a prediction for the next slot function $\p f_{t+1}$, or for its gradient $\p c_{t+1}$, which suffices to solve the caching problem, as we will see.


\textbf{Benchmark}. In such learning problems, it is important to understand the learning objective that our learning algorithm aims to achieve. If we had access to an oracle for the requests $\{q_t\}_{t=1}^T$ (and utility parameters) we could devise the utility-maximizing static caching and routing policy $x^\star=(y^\star, z^\star)$, by solving the following convex optimization problem:
\vspace{-2mm}
\begin{align}
\mathbb P_1:\quad \max_{x}  \,\,\,\,\,\,& \sum_{t=1}^T f_t(x) \label{eq:opt1a}\\
\text{s.t.   }&\,\, z_{nij} \leq y_{nj}\ell_{ij}, \quad i\in\mathcal I, j\in\mathcal J, n\in\mathcal N, \label{eq:opt1b}\\
	& \,\,z\in \mathcal Z, \,\,\,\, y\in\mathcal Y
	\label{eq:opt1c},
\end{align} 
where constraints \eqref{eq:opt1b} ensure the routing decisions for each requested file use only the edge caches that store enough chunks of that file. And let us define the set of constraints $\mathcal X=\big\{\left\{\mathcal Y\times \mathcal X\right\}\cap \{\eqref{eq:opt1b} \}\big\}$, which is compact and convex.

This hypothetical solution $x^\star$ can be designed only with \emph{hindsight} and is the benchmark for evaluating our online learning policy $\pi$. To that end, we use the metric of regret:
\begin{align}\label{eq:regret}
R_T(\pi)= \sup_{ \{f_t\}_{t=1}^T }\left[ \sum_{t=1}^T f_t\big(x^\star\big)-\sum_{t=1}^T f_t\big(x_t\big)\right],
\end{align}
which quantifies the performance gap of $\pi$ from $x^\star$, for any possible sequence of requests or, equivalently, functions $\{f_t\}_t$. Our goal is to find a policy that achieves sublinear regret, $R_T(\pi)\!=\!o(T)$, thus ensuring the average performance gap will diminish as $T$ grows. This policy, similar to other online policies, decides $x_{t+1}$ at the end of each slot $t$ using the previous utility functions $\{f_\tau\}_{\tau=1}^t$ and the next-slot prediction $\tilde f_{t+1}$ which is made available through the rec-sys.

\section{Optimistic Bipartite Caching}\label{sec:bipartite}

Unlike recent caching solutions that rely on Online Gradient Descent (OGD)  \cite{paschos-infocom19} or on the Follow-the-Perturbed-Leader (FTPL) policy \cite{abhishek-sigm20}, our approach draws from the \emph{Follow-The-Regularized-Leader} (FTRL) policy, cf. \cite{mcmahan-survey17}. A key element in our proposal is  the \emph{optimism} emanating from the availability of predictions, namely the content recommendations that are offered to users by the rec-sys in each slot.

\begin{figure}
	\centering
	\includegraphics[width=3.05in]{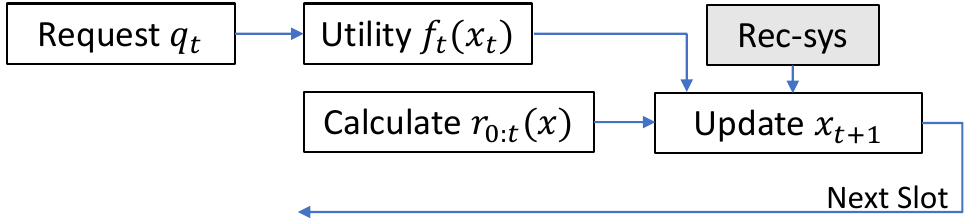} 
 	\vspace{-2mm}	
	\caption{A decision step for OBC. When a request $q_t$ arrives, the file is routed based on the current cache configuration. The caches are updated using the observed utility $f_t(x_t)$ and the new prediction from the recommender.}
	\label{fig:model}
	\vspace{-2mm}
\end{figure}

Let us begin by defining the proximal regularizers\footnote{A proximal regularizer is one that induces a proximal mapping for the objective function; see \cite[Ch. 6.1]{beck-book} for the formal definition.}: 
\begin{align}
r_0(x)=\bm I_{\mathcal X}(x), \quad	r_t(x)=\frac{\sigma_t}{2}\|x-x_t\|^2,\,\,t\geq 1 \label{regula}
\end{align}
where $\|\cdot \|$ is the Euclidean norm, and $\bm I_{\mathcal X}(x)\!=\!0$ if $x\!\in\! \mathcal X$ and $\infty$ otherwise. We apply the following regularizing parameters:
\begin{align}
\sigma_t\!=\!\sigma\! \left(\!\sqrt{ h_{1:t} }\! - \! \sqrt{h_{1:t-1} }\right), \sigma_1\!=\!\sigma\sqrt{h_1},  \ h_t\!=\!\|c_t-\p c_t\|^2  \label{regulb}
\end{align}
where $\sigma\!\geq \!0$, $c_t\!=\! \grad f_t(x_t)$, and we used the shorthand notation $h_{1:t}\!=\!\sum_{i=1}^th_i$ for the aggregate prediction errors during the first $t$ slots.  The basic step of the algorithm is the update: 
\begin{align}
x_{t+1}=\arg\min_{x\in\mathbf R^m}\Big\{ r_{0:t}(x)- (c_{1:t} + \p c_{t+1})^\top x	\Big\},   \label{proxy-step}
\end{align}
which calculates the decision vector based on past observations $c_{1:t}$, the aggregate regularizer $r_{0:t}(x)$ and the prediction $\p c_{t+1}$ (see Fig. \ref{fig:model}). The update employs the negative gradients as it concerns a maximization problem, cf. \cite{mcmahan-survey17}. Henceforth, we refer to \eqref{proxy-step} as the \emph{optimistic} FTRL (OFTRL) update. 

Policy $\pi_{obc}$ is outlined in Algorithm \ref{alg1}. In each iteration, OBC solves a convex optimization problem, \eqref{proxy-step}, involving a projection on the feasible set $\mathcal X$ (via $r_0(x)$). For the latter, one can rely on fast-projection algorithms specialized for caching, e.g., see \cite{paschos-infocom19}; while it is possible to obtain a closed-form solution for the OFTRL update for linear functions.
We quantify next the performance of Algorithm \ref{alg1}. 
\setlength{\textfloatsep}{0pt}
\begin{mdframed}
\begin{theorem}\label{th:regret1}
Algorithm OBC ensures the regret bound:
\begin{align}
R_T\leq 2\sqrt{2(1\!+JC)} \sqrt{\sum_{t=1}^T  \|c_t-\p c_t\|^2 } \notag
\end{align}
\end{theorem}
\end{mdframed}
\begin{proof}
We start from  \cite[Theorem 1]{mohri-aistats2016} which proved that a proximal OFTRL update with regularizer $r_{0:t}(x)$ that is 1-strongly-convex w.r.t. some norm $\|\cdot\|_{(t)}$ yields regret:
\begin{align}
R_T\leq r_{1:T}(x^\star)+\sum_{t=1}^T\|c_t-\p c_t \|_{(t),\star}^2,\,\,\,\,\forall x^\star \in \cal X. \label{mohri-rt}
\end{align}
Now, $r_{1:t}$ is 1-strongly-convex w.r.t. norm $\|x\|_{(t)}=\sqrt{\sigma_{1:t}}\|x\|$ which has dual norm $\|x\|_{(t),\star}=\|x\|/\sqrt \sigma_{1:t}$.
Using the regularization parameter \eqref{regulb}, we get $\sigma_{1:t}=\sigma\sqrt{h_{1:t} }$, and replacing all the above into \eqref{mohri-rt} we get:
\setlength{\textfloatsep}{0pt}
\begin{align}
R_T &\leq \frac{\sigma}{2} \sqrt{h_{1:T}}  D_{\mathcal X} ^2 + \sum_{t=1}^T \frac{h_t}{\sigma\sqrt{h_{1:t}}} \label{eq:reg_int} 
\end{align}
 where we upper bouned each $\|x - x_t\|$ term with the Euclidean diameter of $\mathcal X$ denoted by $D_{\mathcal X}$. Namely $\forall x, x_t\! \in\! \cal X$ holds:
\begin{align*}
&\|x-x_t\|^2=\sum_{n,j}(y_{nj}-y_{nj}^t)^2+\sum_{n,i,j}(z_{nij}-z_{nij}^t)^2 \notag \\
&\stackrel{(a)}{\leq} \sum_{n,j}|y_{nj}-y_{nj}^t|+\sum_{n,i,j}|z_{nij}-z_{nij}^t| \stackrel{(b)}{\leq} 2(JC+1)\triangleq D_{\mathcal X}^2 
\end{align*}
where $(a)$ holds as $y_{nj}, z_{nij}\!\in\![0,1]$; $(b)$ holds by the triangle inequality and definitions of $\mathcal Y, \mathcal Z, \mathcal Q$. Finally, using \cite[Lem. 3.5]{auer-2002} to bound the second regret term in \eqref{eq:reg_int} $\sum_t^T h_t/\sqrt{h_{1:t}}\!\leq\! 2\sqrt{h_{1:T}}$ and setting $\sigma\!=\! 2/D_{\mathcal X}$
 we arrive at the result.
\end{proof}
\setlength{\textfloatsep}{0pt}
\begin{algorithm}[t]
	\SetAlgoRefName{OBC}		
	\nl \textbf{Input}: $\{\ell_{ij}\}_{(i,j)}$; $\{C_j\}_j$; $\mathcal{N}$; $x_1\!\in\!\mathcal X$; $\sigma=2/D_{\mathcal X}$.\\%
	\nl \textbf{Output}: $x_t=(y_t, z_t)$, $\forall t$.\\%
	\nl \For{ $t=1,2,\ldots$  }{
		\nl Route request $q_{t}$  according to configuration $x_t$\\
		\nl Observe system utility $f_t(x_t)$ \\
		\nl Observe the new prediction $\p c_{t+1}$\\
		\nl Update the regularizer $r_{0:t}(x)$ using \eqref{regula}-\eqref{regulb}\\
		\nl Calculate the new  policy $x_{t+1}$ using \eqref{proxy-step}\\
	}
	\caption{{Optimistic Bipartite Caching ($\pi_{obc}$)}}\label{alg1}
	\vspace{-1mm}	
	\setlength{\intextsep}{0pt} 
	
\end{algorithm} 

\textbf{Discussion}. Theorem \eqref{th:regret1} shows that the regret does not depend on the library size $N$ and is also modulated by the quality of the content recommendations; accurate predictions tighten the bound, and in the case of perfect prediction, i.e., when users follow the recommendations, we get a negative regret $R_T \! \leq \! 0, \forall T$, which is much stronger than the sub-linear growth rates in other works \cite{paschos-infocom19,8999784}. On the other hand, for worst-case prediction, it is $\|c_t-\p c_t \|^2 \leq 2w^2$, thus $R_T\leq 4w\sqrt{2(JC+1)}\sqrt{T}=O(\sqrt{T})$; i.e., the regret is at most a constant factor worse than the regret of those policies that do not incorporate predictions\footnote{The factor is $2$ compared to the ``any-time" version of the bound that do not use predictions, and $2\sqrt{2}$ compared to those that assume a known $T$.}, regardless of the predictions' quality. Thus, OBC offers an efficient and safe approach for incorporating predictions in cases where we are uncertain about their accuracy, e.g., either due to the quality of the rec-sys or the behaviour of users.

Another key point is that the \emph{utility parameters might vary with time} as well. Indeed, replacing $w_t=(w_{nij}^t\!\leq \! w, n\!\!\in\!\!\mathcal N, i\!\!\in\!\!\mathcal I, j\!\!\in\!\!\mathcal J)$ in $f_t(x_t)$ does not affect the analysis nor the bound. This is important when the caching system employs a wireless network where the link capacities vary, or when the caching utility changes. Similarly, for edge computing and caching services, the utility of each computation or service might vary substantially across users and time. Parameters $w_t$ can be even unknown to the caching system when $x_t$ is decided, exactly as it is with $q_t$, and they can be predicted either using the rec-sys or other side information (e.g., channel measurements).
\section{Optimistic Caching in Elastic Networks}\label{sec:elastic}

We extend our analysis to \emph{elastic} caching networks where the caches can be resized dynamically. Such architectures are important for two reasons. Firstly, there is a growing number of small-size content providers that implement their services by leasing storage on demand from infrastructure providers \cite{amazon-cache}; and secondly, CDNs often resize their caches responding to the time-varying user needs and operating expenditures \cite{elastic-cdn}.


We introduce the $t$-slot price vector $s_t\!=\!(s_j^t\!\leq \!s, j\!\in\!\mathcal J)$, where $s_j^t$ is the leasing price per unit of storage at cache $j$ in slot $t$, and $s$ its maximum value. In the general case, these prices may change arbitrarily over time, e.g., because the provider has a dynamic pricing scheme or the electricity cost changes \cite{giannakis-elasticJSAC19, jungho-wiopt}; hence the caching system has access only to $s_t$ at each slot $t$. We denote with $B_T$ the budget the system intends to spend during a period of $T$ slots for leasing cache capacity. The objective is to maximize the caching gains while satisfying the constraint:
\begin{align}
\sum_{t=1}^T g_t(x_t)=\sum_{t=1}^T \sum_{j\in\mathcal J}\sum_{n\in\mathcal N} s_j^ty_{nj}^t-B_T\leq 0.
\end{align}
In particular, the new benchmark problem in this case is:
\begin{align}
\mathbb P_2:\quad \max_{x\in\mathcal X} \sum_{t=1}^T f_t(x) \,\,\,\,\,\,\text{s.t.}\,\,\,\,\,\, \eqref{eq:opt1b},\,\,\,\,\, \sum_{t=1}^T g_t(x)\leq 0,
\end{align}
which differs  from $\mathbb P_1$ due to the  leasing constraint. 

Indeed, in this case the regret is defined as:
\begin{align}\label{eq:regret-elastic}
	R_T^{(e)}(\pi)= \sup_{ \{f_t, g_t\}_{t=1}^T }\left[ \sum_{t=1}^T f_t\big(x^\star\big)-\sum_{t=1}^T f_t\big(x_t\big)\right],
\end{align}
where $x^\star\! \in\! \mathcal X_e\triangleq\{ x \in \mathcal X \mid \eqref{eq:opt1b}, g_t(x)\leq 0, \forall t \}$, i.e., $x^\star$ is a feasible point of $\mathbb P_2$ with the additional restriction to satisfy $g_t(x)\!\leq\!0$ in every slot. In the definition of $\mathcal{X}$, $C$ now denotes the maximum leasable space. Learning problems with time-varying constraints are notoriously hard to tackle, see impossibility result in \cite{tsitsiklis}, and hence require such additional restrictions on the selected benchmarks. We refer the reader to \cite{giannakis-TSP17} for a related discussion, and to \cite{paschos-icml, victor}  for more competitive benchmarks. These ideas are directly applicable to our OFTRL framework. For instance, the analysis follows directly for the $K$-slot benchmark of \cite{paschos-icml} where $\sum_{\tau=t}^{t+K}g_t(x^\star)\!\leq\! 0, \forall \tau$, instead of $g_t(x^\star) \!\leq\!0, \forall t$. Finally, apart from $R_T^{(e)}$, we need also to ensure sublinear growth rate for the budget violation:
\begin{equation}
V_T^{(e)}=\sum_{t=1}^T g_t(x_t). \notag
\end{equation}
To tackle this new problem we follow a saddle point analysis, which is new in the context of OFTRL.

We first define a Lagrangian-type function by relaxing the budget constraint and introducing the dual variable $\lambda\geq 0$:
\begin{align}
\mathcal L_t(x,\lambda)\!=\!\frac{\sigma_t}{2}\|x\!-x_t\|^2 \!- f_t(x_t) \! +\lambda g_t(x_t)\!- \frac{\lambda^2}{a_t}. \label{eq:lag}
\end{align}
The last term is a non-proximal regularizer for the dual variable; and we use $a_t\!=\!at^{-\beta}$, where parameter $\beta\!\in\! [0,1)$ can be used to prioritize either $R_T^{(e)}$ or $V_T^{(e)}$. The main ingredients of policy $\pi_{oec}$ are the saddle-point iterations:
\begin{align}
\lambda_{t+1}=\arg\max_{\lambda\geq 0}\left\{-\frac{ \lambda^2}{a_{t+1}} + \lambda\sum_{i=1}^tg_i(x_i) \right\},\label{dual-update}	
\end{align}
\begin{align}
\!\!x_{t+1}\!=\!\arg\min_{x\in\mathbf R^m}\Big\{\!r_{0:t}(x)\!+\!\big( \sum_{i=1}^{t+1}\!\lambda_is_i -c_{1:t}\!-\p c_{t+1} \big)^\top\! x \Big\}
\label{primal-update} 
\end{align}
and its implementation is outlined in Algorithm \ref{alg2}. Note that we use the same regularizer for the primal variables $x_t$, while $\lambda_t$ modulates the caching decisions by serving as a \emph{shadow price} for the average budget expenditure.

\begin{algorithm}[t]
	\SetAlgoRefName{OEC}		
	\nl \textbf{Input}: $\{\ell_{ij}\}_{(i,j)}$, $\{C_j\}_j$, $\mathcal{N}$, $\lambda_1\!=\!0$, $x_1\!\in\!\mathcal X_e$, $a_t\!=\!at^{-\beta}$\\%
	\nl \textbf{Output}: $x_t=(y_t, z_t)$, $\forall t$.\\%
	\nl \For{ $t=1,2,\ldots$  }{
		\nl Route  request $q_{t}$  according to configuration $x_t$\\
		\nl Observe system utility $f_t(x_t)$ and cost $g_t(x_t)$ \\
		\nl Update the budget parameter $\lambda_{t+1}$ using \eqref{dual-update}\\	
		\nl Update the regularizer $r_{0:t}(x)$ using \eqref{regula}-\eqref{regulb}\\
		\nl Observe prediction $\p c_{t+1}$ and price $s_{t+1}$ \\		
		\nl Calculate the new  policy $x_{t+1}$ using \eqref{primal-update} \\
	}
	\caption{{Optimistic Elastic Caching ($\pi_{oec}$)}}\label{alg2}
	\vspace{-1mm}		
\end{algorithm}

The performance of Algorithm OEC is characterized next.
\begin{mdframed}
\vspace{-1.5mm}
\begin{theorem} \label{th:regret-e}
Algorithm OEC ensures the bounds:
\begin{align}
& R_T^{(e)} \leq 2D_{\mathcal X}\sqrt{ \!\sum_{t=1}^T \|c_t\!-\p c_t\|^2 } \!+\frac{a(sJC)^2}{2(1\!-\!\beta)}T^{1-\beta}	 \notag \\
& V_T^{(e)} \leq \sqrt{\!\frac{4 D_{\mathcal X}T^\beta }{a}\sqrt{\!  \sum_{t=1}^T \|c_t\!-\p c_t\|^2  } +\frac{T(sJC)^2}{1\!-\beta} \!-\! \frac{2R_T^{(e)}T^\beta}{a}} \notag 
\end{align}		
\end{theorem}
\vspace{-1.5mm}
\end{mdframed}
\begin{proof}



Observe that the update in \eqref{primal-update} is similar to \eqref{proxy-step} but applied to the Lagrangian in \eqref{eq:lag} instead of just the utility, and the known prices when $x_{t+1}$ is decided represent perfect prediction for $g_t(x)$. Using Theorem \ref{th:regret1} with $c_t \! - \! \ll_ts_t$ instead of $c_t$, and  $\p c_t \! - \!\ll_ts_t$ instead of $\p c_t$, we can write:
\begin{align}
\sum_{t=1}^T\! \Big(f_t(x^\star) - f_t(x_t) +\lambda_tg_t(x_t)-\lambda_tg_t(x^\star)\Big)\!\leq\! 2D_{\mathcal X} \sqrt{ h_{1:T}}, \notag
\end{align}
and rearrange to obtain:
\begin{align}
	R_T^{(e)}\leq 2D_{\mathcal X} \sqrt{ h_{1:T}} + \sum_{t=1}^T \ll_t g_t(x^\star) - \sum_{t=1}^T \ll_t g_t(x_t). \label{r1}
\end{align}
For the dual update \eqref{dual-update}, we can use the non-proximal-FTRL bound \cite[Theorem 1]{mcmahan-survey17} to write:
\begin{align}
\!\!\!\!-\!\sum_{t=1}^T \ll_t g_t(x_t) \!+\! \ll \sum_{t=1}^T g_t(x_t) \!\leq\! \frac{\ll^2}{a_T} \!+\! \frac{1}{2}\sum_{t=1}^T a_tg_t^2(x_t) \label{l1}.
\end{align}
Since $g_t(x^\star)\!\leq\! 0, \forall t$ and combining \eqref{r1}, \eqref{l1} we get:
\begin{align}
\!\!R_T^{(e)}\!\leq 2\!D_{\mathcal X} \sqrt{ h_{1:T} } \!- \ll\! \sum_{t=1}^T g_t(x_t) \!+\! \frac{\ll^2}{a_T} \!+\! \frac{1}{2}\sum_{t=1}^T \!a_tg_t^2(x_t). \label{eqRT1}
\end{align}
Setting $\ll\!=\!0$, using  the identity $\sum_{t=1}^Tat^{-\beta}\!\leq\! aT^{1-\beta}/(1\!-\!\beta)$ and the bound $g_t(x_t)\!\leq\! sJC$, we prove the $R_T^{(e)}$ bound. Using:
\begin{align}
	\frac{a_T}{2} \left[ \sum_{t=1}^T g_t(x_t) \right]^2=\sup_{\lambda\geq 0} \left[\sum_{t=1}^T g_t(x_t)\lambda - \frac{\lambda^2}{2a_T} \right], \notag
\end{align}
we can replace this term to \eqref{eqRT1}  and write:
\begin{align}
	&\frac{a_T}{2}(V_T^{(e)})^2 \leq 2 D_{\mathcal X} \sqrt{ h_{1:T} } +\frac{a(sJC)^2}{2-2\beta}T^{1-\beta} - R_T^{(e)}. \notag
\end{align}
Rearranging and taking the square root yields $V_T^{(e)}$ bound.
\end{proof}

\textbf{Discussion}. The worst-case bounds in Theorem \ref{th:regret-e} arise when the predictions are failing. In that case, we have $\|c_t\!-\!\p c_t\|^2\!\leq\! 2w^2$ and use the bound $-R_T^{(e)}\!\!\!=\!\!\!O(T)$ for the last term of $V_T^{(e)}$, to obtain $R_T^{(e)}\!\!\!=\!\!\!O(T^\kappa)$, with $\kappa\!=\!\max\{1/2, 1\!-\!\beta\}$ while $V_T^{(e)}\!=\!O(T^\phi)$, with $\phi\!=\!\frac{1+\beta}{2}$. Hence, for $\beta\!\!=\!\!1/2$ we achieve the desired sublinear rates $R_T^{(e)}\!\!\!=\!\!O(\sqrt{T}), V_T^{(e)}\!\!\!=\!\!O(T^{3/4})$. However, when the rec-sys manages to predict accurately the user preferences, the performance of $\pi_{oec}$ improves substantially as the first terms in each bound are eliminated. Thus, for bounded $T$, we practically halve the regret and violation bounds.  

It is also interesting to observe the tension between  $V_T^{(e)}$ and $R_T^{(e)}$, which is evident from the $V_T^{(e)}$ bound and the condition $-R_T^{(e)}\!=\! O(T)$. The latter refers to the upper bound of the \emph{negative} regret, thus when it is consistently satisfied (i.e., for all $T$), we obtain an even better result: $\pi_{oec}$ \emph{outperforms} the benchmark. Another likely case is when $-R_T^{(e)}\!=\!O(\sqrt{T})$, i.e., the policy does not outperform the benchmark at a rate larger than $\sqrt{T}$. Then, Theorem \ref{th:regret-e} yields $R_T^{(e)}\!=\!O(T^\kappa)$ with $\kappa\!=\!\max\{1/2, 1-\beta\}$ while $V_T^{(e)}\!=\!O(T^\phi)$ with $\phi\!=\!\max\{1/2, 1/4+\beta/2\}$. Hence, for $\beta\!=\!1/2$ the rates are reduced to $R_T^{(e)}\!=\!O(\sqrt{T}), V_T^{(e)}\!=\!O(\sqrt{T})$.


\section{Caching with non-volatile predictions}

We now introduce a different approach on modeling recommendations as predictions, which, in cases of consistent prediction performance, delivers better regret. Namely, we model the problem of online caching using the \emph{experts model}, see \cite{hazan-book}. The first expert represents a robust learner (referred to as \emph{pessimistic}) and proposes an FTRL-based caching policy without any predictions. The second expert represents an \emph{optimistic} learner and implements a policy that always caches the file predicted to be requested. 
To streamline the presentation, we present the results using a single cache scenario (hence using only $y$ below), but it will become clear that this method can be readily extended to caching networks. 

\begin{figure}[t]
	\centering
	\includegraphics[width=3.245in]{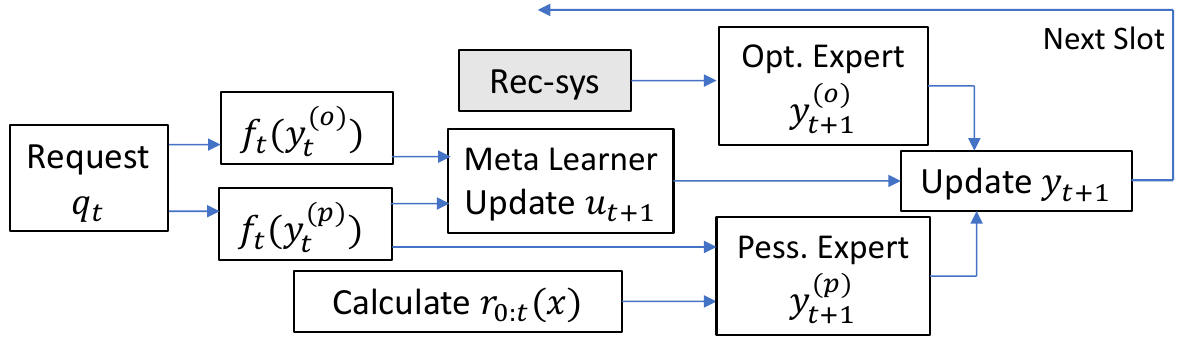} 

	\caption{A decision step for XC. Experts' utilities are used to update the weights $u$. The new caching decisions are then the combination of the experts’ proposals. The optimistic decisions are updated based only on predictions from rec-sys. The pessimistic decisions are updated based only on past requests. 
}
	\label{fig:xc}
	\vspace{-2mm}
\end{figure}

Formally, the pessimistic expert proposes caching actions $\{y^{(p)}_t\}_t$ according to step \eqref{proxy-step}, but with setting $\p c_t\!\! =\!\! 0$ for the regulization parameter $\sigma_t$ in \eqref{regulb}. Its regret w.r.t the optimal-in-hindsight caching configuration $y^\star=\argmax_{y\in\mathcal{Y}} c_{1:T}^\top y$ is denoted with $R_{T}^{(p)}$. On the other hand, the optimistic expert proposes actions $\{y^{(o)}_t\}_t$ according to the linear program:
\begin{align}
    y^{(o)}_{t+1} = \argmax_{y \in \mathcal{Y}}\ \ {\p c_{t+1}}^\top \  y, \label{eq:opt_action}
\end{align}
and we denote its regret with $R_{T}^{(o)}$. The optimistic expert represents a high-risk high-reward policy; $R_{T}^{(o)}$ is linear in the worst case predictions and negative linear for perfect predictions. In contrast, the pessimistic expert is more robust as it is not affected by bad predictions, but guarantees only a sub-linear regret. We aim to have the best of both worlds and design an algorithm that, in the best case, is able to obtain negative regret, while being worse only by a constant factor than the pessimistic expert in the general case.

Unlike $\pi_{obc}$ and $\pi_{ec}$, the predictions are not appended to the FTRL step itself but rather treated independently through the optimistic expert. The challenge is to meta-learn which of the two experts to rely upon. To that end, we will be using Online Gradient Ascent (OGA) to learn how to combine the experts' proposed caching vectors $y^{(p)}_{t}$ and $y^{(o)}_{t}$. The decisions of the meta-learner are then these combination weights $u_t = (u^{(p)}_{t}, u^{(o)}_{t})$, drawn from the 2-dimensional simplex set $\Delta$, (see Fig. \ref{fig:xc}). The weights are learned through the OGA step:

\vspace{-4mm}
\begin{align}
    u_{t+1} = \mathcal{P}_{\Delta} \big\{ u_{t} + \delta_t l_{t}  \big\}, \label{weights_action}
\end{align}
where $\mathcal{P}$ is the projection operator, $\delta_t$ is the OGA learning rate and $l_t=(l_t^{(p)}, l_t^{(o)})$ is the $t$-slot performance vector for the experts, i.e.,  $l_t^{(p)} = {c_t}^\top{y^{(p)}_{t}}$, and $l_t^{(o)} = {c_t}^\top{y^{(o)}_t}$.
The caching decision is the convex combination of experts' proposals:
\begin{align}
    y_{t+1} = u^{(p)}_{t+1}\ y^{(p)}_{t+1} + u^{(o)}_{t+1}\  y^{(o)}_{t+1} \label{eq:comp_action}.
\end{align} Thus, $y_{t+1}$ is still a feasible caching policy. The steps are shown in Algorithm \ref{alg3}, and the following theorem bounds the regret of the \emph{caching decisions} $\{y_t\}_t$. 


\begin{algorithm}[t]
	\SetAlgoRefName{XC}		
	\nl \textbf{Input}: $C$; $y_1\!\in\!\mathcal Y$; $\sigma=2/D_{\mathcal Y}$.\\%
	\nl \textbf{Output}: $y_t$, $\forall t$.\\%
	\nl \For{ $t=1,2,\ldots$  }{
		\nl Serve request $q_{t}$  according to configuration $y_t$\\
		\nl Observe utilities $f_t(y_t^{(p)}), f_t(y_t^{(o)})$\\
		\nl Update $r_{0:t}(x)$ using \eqref{regula}-\eqref{regulb} with $\p c_{t+1}\! = \! 0$\\
		\nl Calculate pessimistic expert's proposal $y^{(p)}_{t+1}$ as in \eqref{proxy-step}\\
		\nl Observe the prediction $\p q_{t+1}$, and calculate $\p c_{t+1}$\\
		\nl Calculate optimistic expert's proposal $y^{(o)}_{t+1}$ using \eqref{eq:opt_action}\\
		\nl Calculate the new weights $u_{t+1}$ using  \eqref{weights_action}\\
		\nl Calculate the new  policy $y_{t+1}$ using \eqref{eq:comp_action}
	}
	\caption{{Experts Caching ($\pi_{xc}$)}}\label{alg3}
	\vspace{-1mm}	
\end{algorithm} 



\begin{mdframed}
\begin{theorem}\label{th:regret3}
Algorithm XC ensures the bound $R^{(xc)}_T$=
\begin{align}
     \! \sum_{t=1}^T \! {c_t}^\top \! (y^\star \!-\! y_t) \leq 2w\sqrt{2T}+\! A, \  A\! \in \! [-wT, 2w\sqrt{2CT}] \notag
\end{align}
\end{theorem}
\end{mdframed}

\begin{proof}
First, we relate the regret of the combined caching decisions to that of the expert selection,

\vspace{-5mm}
\begin{align} \!
    R^{(xc)}_T \! &=  \! \sum_{t=1}^T \! {c_t}^\top{y^\star} \! - \! {c_t}^\top(u^{(p)}_{t} y^{(p)}_t \! + \! u^{(o)}_{t} y^{(o)}_t) \!=\! \sum_{t=1}^ T {c_t}\!^\top\!{y^\star}\! - \! {l_t}\!^\top\!{u_t} \notag
    \\
    &= \sum_{t=1}^ T {c_t}^\top{y^\star} -  {l_t}^\top{u^\star} + {l_t}^\top{u^\star} - {l_t}^\top{u_t}  \notag \\
    &=    R_T^{(u)} + \min \left\{ R_{T}^{(p)} , R_{T}^{(o)} \right\} \label{eq:comp_reg},
\end{align}
where $R_T^{(u)}$ is the regret for the \emph{expert selection weights $u$}: $R_T^{(u)} = \sum_{t=1}^T  {l_t}^\top{u^\star} - {l_t}^\top{u_t} $. \eqref{eq:comp_reg} holds because $u^\star \!\! =\! \argmax_{u\in\Delta}\ {l_{1:t}}\!\!^\top \! {u} \! =\max\{l_t^{(o)}, l_t^{(o)}\} $ Thus, we have that
\begin{align}
l_{1:t}^\top\ u^\star = max\bigg\{ \sum_{t=1}^T c_{t}^\top y^{(p)}_t, \sum_{t=1}^T c_{t}^\top y^{(o)}_t\bigg\}.
\end{align}
Now, we write the expressions for the terms in \eqref{eq:comp_reg}. $R_{T}^{(p)}$ can be bounded in the same manner as Theorem $1$ with prediction vectors $\p c_t =0$, and substituting an upper bound $w$ for $\|c_t\|$:
\begin{align}
   R_{T}^{(p)} &\leq  2w \mathcal{D}_{\mathcal{Y}} \sqrt{T}
    \leq  2w \sqrt{2CT}. \label{eq:regbound_rob}
\end{align}
$R_{T}^{(o)}$ is hard to calculate as it depends on both, prediction $\{\tilde{c}_t\}_t$, and the relationship between $c_{1:t}$ and $c_t$. However, we can easily deduce lower and upper bounds. Since $c_t$ and $\tilde{c}_t$ represent the utility of one request to a file, each term of the optimistic regret can be maximally $w$. Hence, we have that $R_{T}^{(o)}\! \in\! [-wT, wT]$, and:
\begin{align}
    \min \left\{ R_{T}^{(p)} , R_{T}^{(o)} \right\} \in [-wT, 2w\sqrt{2CT}] \label{eq:minrange}.
\end{align}
For $R_T^{(u)}$, we use the OGA bound \cite[Thm. 4.14]{orabona2021modern} tailored to the simplex decision set. Using $\eta_t^{(u)}\! =\! \frac{1}{w\sqrt{t}}$, and $\|l_t\| \! \leq \! \sqrt{2}w$:
\vspace{-3mm}
\begin{align}
    R_T^{(u)} & \leq 2w\sqrt{2T} \label{eq:weights_bound}
\end{align}
Substituting \eqref{eq:minrange} and \eqref{eq:weights_bound} in \eqref{eq:comp_reg} gives the bound.
\end{proof}

\textbf{Discussion}. The regret in Theorem \ref{th:regret3} can now be \emph{strictly} negative for perfect predictions, which is tighter than $\pi_{obc}$. In general, however, the regret bound can be worse than that of $\pi_{obc}$. Namely, $R_T^{(xc)}$ is bounded by $2w\sqrt{2T} + 2\sqrt{2}w\sqrt{CT}$, while in the single cache case  $R_T$ is bounded by $2\sqrt{2C}\sqrt{\sum_t\|c_t-\tilde c_t\|^2}$, which might be better or worse that $R_T^{(xc)}$, depending on the number of steps with accurate predictions. In all cases, $R_T^{(xc)}=O(\sqrt{T})$. An additional point to highlight is that the first term, $2w\sqrt{2T}$, is a worst-case bound for finding the best expert, i.e.,  $R_{T}^{(u)}$. In request sequences where the best expert is easily identifiable, e.g., due to consistent predictions which make $l_t$ similar, it would be a loose upper bound and its actual value is constant (negligible) compared to the second term $\min \left\{ R_{T}^{(p)} , R_{T}^{(o)} \right\}$. This second term is the regret of the best expert, and falls in a range that depends on predictions' quality at each step. Thus, if the optimistic expert is better than the best-in-hindsight solution, this $\min$ term will be negative for some request sequences.



\section{Performance evaluation}\label{sec:evaluation}
\begin{figure}[t]
\centering
\includegraphics[width=0.495\textwidth]{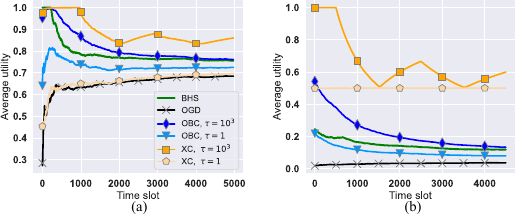}
\vspace{-7mm}
\caption{Utility in the single cache model under different prediction quality levels in (a) Zipf requests with $\zeta=1.2$, (b) YouTube request traces \cite{zink2008watch}.}
\label{fig:single_cache}
\vspace{-2mm}
\end{figure}
We evaluate $\pi_{obc}$, $\pi_{oec}$ and $\pi_{xc}$ under different request patterns and predictions modes; and we benchmark them against $x^\star$ and the OGD policy \cite{paschos-infocom19} that outperforms other state-of-the-art policies \cite{giovanidis-mLRU, leonardi-implicit}. We observe that when reasonable predictions are available, the proposed policies have an advantage, and under noisy predictions, they still reduce the regret at the same rate with OGD, as proven in the Theorems. First, we compare $\pi_{obc}$ and $\pi_{xc}$ against OGD \cite{paschos-infocom19} in the single cache case. We then study $\pi_{obc}$ for the bipartite model and $\pi_{oec}$ with the presence of budget constraints. We consider two requests scenarios, stationary Zipf requests (with parameter $\zeta\!=\!1.2$) and an actual trace from the dataset in \cite{zink2008watch}. Predictions alternate between accurate and adversarial (i.e., requesting the recommended file vs. any other file, respectively), for $\tau$ time step in each mode. While low values of $\tau$ represent an unstable performance, the high value of $\tau$ is meant to approximate the consistent performance of practical rec-sys. We also experiment with random accuracies where at each $t$, the prediction is accurate with probability $\rho$.

\textbf{Single Cache Scenarios}.
We set $w\!=\!1$ to study the cache hit rate scenario. Fig. \ref{fig:single_cache}.a shows the performance of $\pi_{obc}$ and $\pi_{xc}$ for a library of $N\!=\!10^4$ files and cache size $C\!=\!100$. At each slot, we plot the attained average utility, $\frac{1}{T}\sum_{t=1}^T f_t(x_t)$, for each policy and the best static cache configuration \emph{until that slot}, i.e., we find the best in hindsight\footnote{unlike \cite{paschos-infocom19} that calculates $x^\star$ for the largest $t$, we use $x_t^\star$ for each $t$. Thus, the gap between any policy and BHS at $t$ is the policy's average regret $R_t/t$.} for each $t$. 

In the simulated requests case (Fig. \ref{fig:single_cache}.a), $\pi_{obc}$ achieves negative regret through the experiment for $\tau\!\!=\!\!10^{3}$ and a regret that is $57.1\%$ better than that of the OGD for $\tau\!\!=\!\!1$. Such an advantage for the former is due to having more time steps with accurate predictions. $\pi_{xc}$ also maintains negative regret that even outperforms $\pi_{obc}$ when $\tau\!=\!10^{3}$. This is because the stable performance of experts allows the policy to efficiently find the best expert and stick with it within each time window. However, a value of $\tau\!\!=\!\!1$ induces frequent switching between the two experts in $\pi_{xc}$: the performance of the optimistic expert alternate between $0$ and $1$, while that of pessimistic expert is in the range ($0.6, 0.7$). Hence, $\pi_{xc}$ is inclined to place some weight on the opt. expert at one step, only to retract and suffer a greater loss at the following one had it stayed with the full weight on the pess. expert. Due to the additional regret caused by such frequent switching, $\pi_{obc}$ performs better when $\tau=1$. 

For the trace used in Fig. \ref{fig:single_cache}.b, $\pi_{obc}$ maintains the advantage over OGD in both prediction modes. Regarding $\pi_{xc}$, the alternation of the performance of the opt. expert (when $\!\tau\!\!=\!\!1$) no longer induces a switching between the experts since even when the opt. expert performs poorly (gets $0$ reward), there is a high chance, especially initially, that the pess. perform similarly\footnote{The poor performance of the pess. expert here is due to more uniform, unpredictable, request vectors}. Hence, finding that the opt. expert is better is still easy (due to differences in their utility). Thus, in this trace, $\pi_{xc}$ performs well with both $\tau$ values.


\begin{figure}
\centering
	\includegraphics[width=0.46\textwidth]{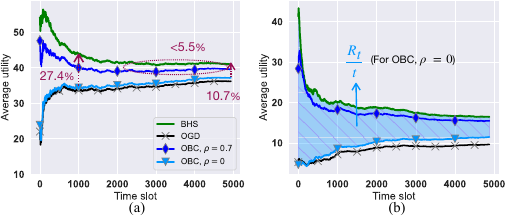}
	\vspace{-3mm}
	\caption{Attained utility in the bipartite model under different prediction quality levels in (a) Zipf requests with $\zeta=1.2$, (b) YouTube request traces \cite{zink2008watch}.}
	\label{fig:bi_unconst}
	\vspace{-2mm}
\end{figure}
\textbf{Bipartite Networks}. We consider next a bipartite graph with $3$ caches and $4$ user locations, where the first two locations are connected with caches $1$ and $2$, and the rest are connected to caches $2$ and $3$. The utility vector is $w_n\!=\!(1, 2, 100), \forall i,j$, thus an efficient policy places popular files on cache $3$. This is the setup used in \cite{paschos-infocom19} that we adopt here to make a fair comparison. For the stationary scenario, we consider a library of $N\!=\!500$ files and $C\!=\!50$. For the traces scenario, files with at least $10$ requests are considered, forming a library of $N\!=\!456$ files, and we keep $C\!=\!50$. In this experiment, we assume that at each time step, the user follows the recommendation with probability $\rho$. The location of each request is selected uniformly at random. Similar to the single-cache case, we plot the average utility of the online policies and the best static configuration \emph{until each $t$}. 

Scenario $1$ in Fig. \ref{fig:bi_unconst}.a shows the effect of good predictions as OBC maintains utility within $5.32\%$ of BHS's utility after $t\!\!=\!\!2.5k$. Even when the recommendations are not followed, OBC preserves the sublinear regret, achieving a gap of $27.4\%$ and $10.36\%$ for  $t\!\!=\!\!1k$ and $t\!\!=\!\!5k$, respectively. Akin patterns appear in the second scenario (Fig. \ref{fig:bi_unconst}.b) but with lower utilities across all policies due to the more spread requests. Recall that the area between a policy and BHS is the average regret.







\begin{figure}
\centering
	\includegraphics[width=0.5\textwidth]{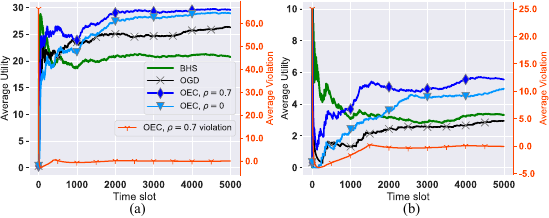}
	\vspace{-7mm}
	\caption{Attained utility and constraints violations for OEC, OGD with (a): Zipf requests with $\zeta=1.5$ and (b): YouTube request traces \cite{zink2008watch}.}
    \vspace{-2mm}
	\label{fig:bi_const_st}
\end{figure}
Next, we consider the case of budget constraint and evaluate $\pi_{oec}$ for scenario 1, Fig. \ref{fig:bi_const_st}.a,  and scenario 2, Fig. \ref{fig:bi_const_st}.b. The prices at each slot are generated uniformly at random in the normalized range $[0, 1]$, and the available budget is generated randomly $b_t = \mathcal{N}(0.5, 0.05) \times 10$, i.e., enough for approximately $10$ files. Such tight budgets magnify the role of dual variables and allow us to test the constraint satisfaction. The benchmark $x^\star$ is computed once for the \emph{full time horizon}, and its utility is plotted for each $t$. In both scenarios, we note that the constraint violation for all policies is approximately similar, fluctuating during the first few slots and then stabilizing at zero. Hence, we plot it for one case.

Concluding, we find that $\pi_{oec}$ can even outperform the benchmark since it is allowed to violate the constraints at some time slots, provided that the constraints are eventually satisfied, which occurs either due to strict satisfaction or due to having an ample subsidy at some slots. Moreover, in the first scenario (Fig.\ref{fig:bi_const_st}.a), the good predictions enable OEC to outperform $x^\star$ by $42.5\%$ after observing all requests ($T\!\!=\!\!5K$). OGD, and OEC with noisy predictions attain utility units improvement of $26.5\%$, $39.3\%$, respectively, over the BHS. In the second scenario  (Fig.\ref{fig:bi_const_st}.b) , the good forecast enables a utility gain of $67.1\%$ compared to, $-11.3\%$, and $49.7\%$ for OGD and OEC with  noisy prediction, respectively. The algorithms scale for very large libraries $\mathcal N$, and the only bottleneck is finding $x^\star$ which involves the horizon $T$, see also \cite{paschos-infocom19, abhishek-sigm20}; this is not required in real systems. The code for the presented policies and experiments is available at \cite{code}.



\section{Conclusions}\label{sec:conclusions}

The problem of online caching is timely with applications that extend beyond content delivery to edge computing \cite{paschos-jsac}. This work proposes a new suite of caching policies that leverage predictions obtained from content-viewing recommendations to achieve negative regret w.r.t to an ideal (unknown) benchmark. As recommender systems permeate online content platforms, such policies can play an essential role in optimizing the caching efficacy. We identified and built upon this new connection. The framework is scalable and robust to the quality of recommendations, improves previously known caching regret bounds \cite{paschos-infocom19, stratis-2020, abhishek-sigm20}, and opens new directions. Among them, the design of optimistic policies for uncoded caching is perhaps the most promising.

\section{Acknowledgment}
This publication has emanated from research conducted with
the financial support of the European Commission through Grant No. 101017109 (DAEMON).

\bibliography{./references_short}
\bibliographystyle{IEEEtran}
\end{document}